\documentclass[11pt,onecolumn]{article}
\usepackage{setspace,dcolumn}
\usepackage{subfig}
\usepackage{hyperref}
\usepackage{cite}
\usepackage{graphicx,psfrag,epsfig,graphbox}
\usepackage[font=small,format=plain,labelfont=bf,justification=raggedright,singlelinecheck=false]{caption}
\usepackage{amsmath,amsfonts,amssymb,amsthm,bm,upgreek}
\allowdisplaybreaks[2]
\usepackage{geometry}
\usepackage[mathscr]{eucal}
\usepackage{esvect}
\usepackage{booktabs,threeparttable}
\usepackage{indentfirst}
\usepackage[normalem]{ulem}
\usepackage{lscape}
\usepackage{physics}
\usepackage{subfiles}
\usepackage{multirow}
\usepackage{tikz}
\usepackage{tikz-cd}
\usepackage{epstopdf}

\usepackage{extarrows}

\usepackage{color}
\definecolor{darkgreen}{RGB}{40,150,60}
\definecolor{violet}{RGB}{140,50,230}
\definecolor{orange}{RGB}{230,150,0}

\hypersetup{colorlinks=true,linkcolor=blue,anchorcolor=blue,citecolor=blue,urlcolor=blue}
\numberwithin{equation}{section}

\title{Geodesics in Carrollian Reissner-Nordstr\"om black holes}
\author{Bin Chen$^{1,2,3,4}$, Haowei Sun$^3$, Jie Xu$^3$}

\begin{document}
\maketitle
\begin{center}
	{\it

        $^{1}$Institute of Fundamental Physics and Quantum Technology, Ningbo University, Ningbo, Zhejiang 315211, China\\\vspace{4mm}
 
        $^{2}$ School of Physical Science and Technology, Ningbo University, Ningbo, Zhejiang 315211, China\\\vspace{4mm}

		$^{3}$School of Physics, Peking University, No.5 Yiheyuan Rd, Beijing 100871, P.~R.~China\\
		\vspace{2mm}
		
		$^{4}$Center for High Energy Physics, Peking University, No.5 Yiheyuan Rd, Beijing 100871, P.~R.~China\\
	}
	\vspace{10mm}
\end{center}

\begin{abstract}
    \vspace{5mm}
    \begin{spacing}{1.5}
       In this work, we study the geodesics in different types of Carrollian RN (Reissner-Nordstr\"om) black holes, considering the motions of both neutral and charged particles. We use the geodesic equations in the weak Carrollian structure and analyze the corresponding trajectories projected onto the absolute space, and find that the geodesics are well-defined. In particular, we examine the electric-electric and magnetic-electric limit of the RN black hole, focusing on their geodesic structures. We find that the global structures of the usual RN black holes get squeezed under the ultra-relativistic limit. More precisely, the nonextreme magnetic-electric RN spacetime has two different asymptotic flat patches while the extreme black hole spacetime consists of only one patch. For the magnetic-electric RN spacetime, the Carrollian extremal surfaces (CESs) divide the spacetime into several geodesically complete regions, and the geodesics can only travel in one of these regions. For the charged particles, we extend the analysis by considering their interactions with the electromagnetic field in the Carrollian RN spacetimes and find that their trajectories are significantly different from the neutral geodesics.
      
    \end{spacing}
\end{abstract}
\newpage

\setcounter{tocdepth}{2}
\tableofcontents

\section{Introduction}\label{sec:Introduction}

    In the early 20th century, Einstein's theory of general relativity revolutionized people's understanding of gravity. Shortly after, in 1916 Karl Schwarzschild provided the first exact solution of Einstein's equations. Hans Reissner and Gunnar Nordström later extended this solution to include the Maxwell theory, resulting in the Reissner-Nordström (RN) metric describing the spacetime geometry around a non-rotating electrically/magnetically charged spherically body. Although real black holes formed in nature are not expected to carry significant electric charge, the RN solution remains crucial for theoretical study  and helps in understanding how electromagnetic interaction affects spacetime geometry.

    Einstein's theory of general relativity, with its elegant description of gravity in terms of curved spacetime, has become the cornerstone of modern physics and has been extensively tested and validated. However, it is not the only framework to understand gravity. The classical Newtonian view of spacetime could evolve into a new form, the Newton-Cartan gravity, which generalizes the concept of Newtonian gravity by incorporating additional geometric structures that gauges the non-relativistic Galilean symmetry or Bargmann symmetry \cite{Cartan:1924yea,Andringa:2010it,Bennett:2021dbg,Hartong:2022lsy}. This approach bridges Newtonian and relativistic views and offers new insights into how gravity can be described in a non-relativistic context.  
    
    In contrast to the Newtonian limit, the Carrollian framework introduces a totally different ultra-relativistic perspective to include the notion of absolute space. The Carrollian gravity, as a counterpart to the Newton-Cartan gravity, explores the extreme limits of spacetime dynamics where the speed of light approaches zero \cite{Dautcourt:1997hb, Bekaert:2015xua, Hartong:2015xda, Bergshoeff:2017btm, Hansen:2021fxi, Henneaux:2021yzg, Figueroa-OFarrill:2022mcy, Bergshoeff:2023rkk, Ecker:2023uwm, deBoer:2023fnj, Tadros:2023teq,Tadros:2024qlo}. The Carrollian symmetry was first proposed by L\'evy-Leblond in 1965\cite{Levy-Leblond:1965} and independently by Sen Gupta \cite{SenGupta:1966qer}. The Carrollian symmetry has been found in various physical systems, such as the plane-gravitational wave \cite{Souriau:1973,Duval:2017els}, near-horizon region of black hole \cite{Penna:2018gfx,Donnay:2019jiz,Freidel:2022vjq,Redondo-Yuste:2022czg}, cosmology \cite{deBoer:2021jej}, fracton \cite{Casalbuoni:2021fel, Pena-Benitez:2021ipo, Bidussi:2021nmp, Jain:2021ibh, Figueroa-OFarrill:2023vbj, Figueroa-OFarrill:2023qty, Armas:2023dcz}, and in flat holography \cite{Donnay:2022aba,Donnay:2022wvx,Bagchi:2022emh,Chen:2023naw}. It has been incorporated into the string theory \cite{Bagchi:2020fpr, Bagchi:2021rfw, Chen:2023esw,Gomis:2023eav} and the field theories \cite{Basu:2018dub, Barducci:2018thr, Bagchi:2019clu, Bagchi:2019xfx, Chen:2020vvn, Banerjee:2020qjj, Chen:2021xkw, Campoleoni:2021blr, Henneaux:2021yzg, Hao:2021urq,Chen:2022jhx, Chen:2022cpx, Hao:2022xhq, Yu:2022bcp,Banerjee:2022ocj, Bagchi:2022eui, Chen:2023pqf, deBoer:2023fnj, Islam:2023rnc,Ciambelli:2023xqk,Islam:2023iju, Chen:2024voz, Cotler:2024xhb,Banerjee:2024hvb},  One can find more other works in \cite{Aggarwal:2024yxy,Bagchi:2024unl,He:2024yzx,Bergshoeff:2024ytq,Kasikci:2023zdn,Nguyen:2023miw,Chen:2024vho,OConnor:2024rku,Mehra:2024zqv,GUPTA:2024tcd,Correa:2024qej,Ruzziconi:2024zkr} and references therein.

    In \cite{Hansen:2021fxi}, the two limits of Einstein's gravity were obtained from the leading-order and the next-to-leading-order  expansions in $c$ and told us that the different limits of solutions of Einstein's gravity become the solutions of corresponding Carrollian gravity. As expected,  the Carrollian limits of the RN solution may give solutions to the Carrollian gravity coupled to Carrollian electromagnetism. The  Carrollian RN black hole was introduced first in \cite{Ecker:2023uwm} and further analyzed in \cite{Ciambelli:2023tzb,Aggarwal:2024yxy,Tadros:2024bev}. In relativistic case, even though the RN black hole share some important features with the Schwarzschild black hole, including the isometry and asymptotical flatness, it is very different from the Schwarzschild black hole in many aspects. In particular, there are two horizons  in the RN black hole.  More importantly, the global causal structure of the RN black hole is quite different, for example, there exist infinite asymptotic regions. Moreover, there exists electromagnetic force between charged particle and the RN black hole, besides the usual gravitational force. Therefore, it is essential to study the Carrollian RN black hole, in order to have a better understanding of the Carrollian gravity and Carrollian electromagnetism.

    Geodesic plays an indispensable role in studying gravity and spacetime. On one hand, it tells us  how the probe moves under the influence of gravity without any additional forces acting on it. On the other hand, the geodesics and their maximal extensions provide important information about the structure of spacetime. Recently, the geodesics in the Carrollian Schwarzschild black holes have been studied in \cite{Ciambelli:2023tzb, Tadros:2024bev}. In this work, we would like to study the geodesics in Carrollian RN black hole, which is more challenging and may shed new light on the Carrollian gravity and ultra-relativistic physics. 
    
    This  work intends to focus on the Carrollian RN black holes, but within a slightly different geometric framework from the one in  \cite{Ciambelli:2023tzb}, by using the geodesic equations given in \cite{deBoer:2023fnj}, with the reasons being explained later. We will discuss the geodesics in  different types  Carrollian RN black holes, both for neutral and charged particles\footnote{Strictly speaking, due to the presence of electromagnetic force, the motion of a charged particle does not follow a geodesic. In other words, the standard geodesic equation gets modified. In this work, we loosely refer to the solutions of the modified geodesic equation as geodesics as well. }. We will begin by reviewing the foundational concepts of Carrollian geometry and geodesics, then we classify the Carrollian limits of the RN black hole, followed by a detailed analysis of the resulting geodesic equations. Through this exploration, we try to shed light on the distinctive characteristics of the Carrollian RN black hole and  broaden our  understanding of black hole physics in modified spacetime symmetries.

\section{Carrollian geometry and geodesics}\label{sec:CarrollianSymmety}
    To study the geodesics of the Carrollian RN black holes, we need to know the  metrics and the geodesic equations in Carrollian spacetime. In this section, we aim to give a brief review on the Carrollian geometry and Carrollian gravity, and show how to take different Carrollian limits of the relativistic RN black holes to get the Carrollian RN black hole metrics. \par

    The ultra-relativistic (Carrollian) limit is achieved by taking the speed of light to zero, $c \to 0$. Under the Carrollian limit, the Lorentzian boosts transform into the Carrollian boosts
   \begin{equation}\label{eq:GlobalBoost}
        \left\{
        \begin{aligned}
            t&\to t- \mathbf{b} \cdot \mathbf{x},\\
            \mathbf{x}&\to \mathbf{x},\\
        \end{aligned}
        \right.
    \end{equation} such that the Poincaré group transforms into the Carrollian group.

    Einstein's general relativity (GR) is formulated using (pseudo-)Riemannian geometry, and the relativistic gravity is described in terms of a curved spacetime with a (pseudo-)Riemannian metric governed by the Einstein equations. Accordingly, the non-relativistic and ultra-relativistic theories of gravity can be described in the framework of the Newton-Cartan geometry and the Carrollian geometry, respectively.  On the other hand,  
    general relativity can be formulated as a gauge theory of the Poincaré algebra \cite{Utiyama:1956sy,Sciama:1962,Kibble:1961ba} in terms of the spin connection and the tetrads. Similarly,  the Carrollian gravity has been established intrinsically as a gauge theory of the Carrollian algebra \cite{Duval:2014uoa,Ciambelli:2019lap,Herfray:2021qmp,Figueroa-OFarrill:2022mcy}.

    The Carrollian geometry is defined by a Carrollian structure, which is composed of a triplet $(\mathcal{C}, h_{\mu\nu}, n^\mu)$, with $h_{\mu\nu}$ a degenerate metric and $n^\mu$ the temporal vector in its kernel. This naturally gives a fiber bundle structure on $\mathcal{C}$, by treating the integral curves of $n^\mu$ as fibers. In flat spacetime, relativistic physics is defined on a Minkowski spacetime with the metric $ds^2 = -d(x^0)^2 + d(x^i)^2$, invariant under the Poincaré group. Analogous to the fact that  the Lorentz group is the isometry group of flat Minkowski spacetime, the Carrollian group is the isometry group of the flat Carrollian structure defined by
    \begin{equation}
        h_{\mu\nu} = \delta_{ij} dx^i dx^j, \quad n^\mu = \partial_0.
    \end{equation}
But different  from the fact that relativistic metric has a unique inverse, if the Carrollian structure $(n^\mu, h_{\mu\nu})$ has an inverse represented by $(\tau_\mu, \gamma^{\mu\nu})$
    \begin{equation}
        \begin{aligned}
            &n^\mu h_{\mu\nu} = \tau_{\mu}\gamma^{\mu\nu}=0, \quad n^\mu\tau_{\mu}=1,\\
            &n^\mu\tau_{\nu} + \gamma^{\mu\rho}h_{\rho\nu} = \delta^{\mu}_{\ \nu},\\
        \end{aligned}
    \end{equation} 
    any  structure differs by a coordinate-dependent local vector $b_\mu(x)$ is also  {\it a bona fide} inverse
     \begin{equation}\label{eq:LocalBoost}
        \tau_\mu \to \tau_\mu + b_\mu, \quad \gamma^{\mu\nu} \to \gamma^{\mu\nu} + 2 b^{(\mu}n^{\nu)} \qquad \text{where} \qquad n^\mu b_\mu=0, \quad b^\mu = \gamma^{\mu\nu}b_\nu.
    \end{equation} This should be identified as a gauge symmetry of the Carrollian theory, known as the local boost symmetry. Any physical quantity must be invariant under these local transformations generated by $b_\mu(x)$. In other words, the Carrollian physics should be independent of the choice of $\tau_\mu$,  indicating the non-uniqueness of the inverse Carrollian structure and the importance of gauge symmetry in non-Riemannian geometries. This gauge symmetry, distinct from the global boost transformation, is a key feature of Carrollian geometry. Similar feature  has been found in the Newton-Cartan geometry, where it is known as the Milne boost symmetry. For more details on Newton-Cartan geometry and gravity, see \cite{Bennett:2021dbg,Hartong:2022lsy}.

    In GR, a geodesic is the trajectory of a probe particle in the curved spacetime, whose spacetime interval between two events takes extremum value. The geodesic equation follows either from the principle of parallel transport of the tangent vector
    \begin{equation}
        \nabla_{\hat  t} \hat t=0, \qquad \hat t = \frac{d x^\mu}{d \lambda},
    \end{equation} or from the principle of least action of a massive particle 
    \begin{equation}
        I=-m\int d\lambda \sqrt{-g_{\mu \nu} \frac{d x^\mu}{d \lambda} \frac{d x^\nu}{d \lambda}}.\label{firstorderaction}
    \end{equation} Specially, to deal with both massive and massless cases, one can introduce the auxiliary  veilbein field $e(\lambda)$ and modify the action to be
    \begin{equation}
        I=\frac{1}{2}\int d\lambda \left( e^{-1}(\lambda) g_{\mu \nu} \frac{d x^\mu}{d \lambda} \frac{d x^\nu}{d \lambda} -e(\lambda) m^2\right).
    \end{equation} In GR, these two definitions are consistent with each other, provided that the parallel transport is  endowed with the torsionless, metric compatible affine connection. However,  when we turn to the Carrollian case, we will run into difficulty in generalizing the concept of metric compatible connection. As shown in \cite{Figueroa-OFarrill:2020gpr,Hartong:2015xda,Hansen:2021fxi}, a torsionless Carrollian metric-compatible connection does not always exist, and even if it exists it is not unique. This is known as the intrinsic torsion of a Carrollian structure. Consequently,   the geodesic cannot be simply defined by means of parallel transport, and the correct way is to  start from the geodesic action by using the Carrollian structure. In \cite{deBoer:2023fnj}, it has been shown that the action can be directly obtained by replacing the background metric $g_{\mu \nu}$ in  \eqref{firstorderaction} with the Carrollian degenerate metric $h_{\mu \nu}$ and changing the sign,
    \begin{equation}
        I=\frac{1}{2}\int d\lambda \left( e^{-1}(\lambda) h_{\mu \nu} \frac{d x^\mu}{d \lambda} \frac{d x^\nu}{d \lambda} + e(\lambda) m^2\right).
    \end{equation} This is the only known geodesic action that solely  relies on the Carrollian structure while preserves both diffeomorphism and local boost symmetries. We can take an affine parameter $\tau$ such that
    \begin{equation}
        h_{\mu \nu} \frac{d x^\mu}{d \tau} \frac{d x^\nu}{d \tau} = \epsilon = 0,1 .
    \end{equation}
    Here for the massive cases $\tau = \int_A^B d\lambda \sqrt{h_{\mu \nu} \frac{d x^\mu}{d \lambda} \frac{d x^\nu}{d \lambda}}$ is the Carrollian `proper time'\footnote{This could be a misnomer because the `massive' geodesics are actually the limits of relativistic space-like tachyonic geodesics, so the physical meaning of $\tau$ could be more close to measurement of intrinsic space-like distance rather than time interval. It was argued in \cite{deBoer:2021jej,deBoer:2023fnj,Casalbuoni:2023bbh} that such tachyonic particles are important in Carrollian physics.}.  Since $h_{\mu \nu}$ is degenerate but positive semi-definite, there are only two types of Carrollian geodesics  rather than three ones, namely null one and massive one. The resulting geodesic equation is of the form
    \begin{equation}
        h_{\sigma\lambda} \frac{d^2 x^\lambda}{d\tau^2} + \frac{1}{2}  \left( \frac{\partial h_{\sigma\mu}}{\partial x^\nu} + \frac{\partial h_{\sigma\nu}}{\partial x^\mu} - \frac{\partial h_{\mu\nu}}{\partial x^\sigma} \right) \frac{dx^\mu}{d\tau} \frac{dx^\nu}{d\tau} = 0.
    \end{equation} As shown in \cite{deBoer:2023fnj}, the geodesic equation matches the $c\to 0$ expansion and looks quite similar to the relativistic geodesic equation
    \begin{equation}
        \frac{d^2 x^\lambda}{d\tau^2} + \Gamma^\lambda_{\mu\nu} \frac{dx^\mu}{d\tau} \frac{dx^\nu}{d\tau} = 0,\qquad \Gamma^\lambda_{\mu\nu} = g^{\lambda\sigma} \left( \frac{\partial g_{\sigma\mu}}{\partial x^\nu} + \frac{\partial g_{\sigma\nu}}{\partial x^\mu} - \frac{\partial g_{\mu\nu}}{\partial x^\sigma} \right).
    \end{equation} But it is  worth noticing that this  form of Carrollian geodesic equation has differences in the placement of indices compared to the common form of the relativistic geodesic equation because we do not have an inverse metric to lift the indices. In particular, for null geodesics, the geodesic equation becomes
    \begin{equation}\label{eq:nullgeo}
        h_{\mu\nu} \frac{d x^\nu}{d \tau} = 0,
    \end{equation} 
    which represents the famous not-moving Carrollian particles.

    The disadvantage of this form of Carrollian geodesics is that sometimes it cannot produce full geodesic dynamics. For example, if $n^i = h_{0\mu}= 0$ and $h_{\mu\nu}$ is static, the temporal component $x^0(\tau)$ of these equations with respect to the proper time cannot be determined, leading to ambiguity in the geodesics. In \cite{Ciambelli:2023tzb}, this problem was addressed by adding an extra term into the action, but the cost is that the action there used a ruled Carrollian structure\footnote{A ruled Carrollian structure \cite{Ciambelli:2023mir} or a stretched Carrollian structure \cite{Freidel:2024emv} is a modified version of a Carrollian structure by adding an Ehresmann connection to the triplet of the Carrollian structure. The additional Ehresmann connection is an 1-form $\tau_\mu$ dual to the vector $n^\mu$. It is expected that the ruled Carrollian structure captured the intrinsic geometry of a null surface \cite{Bekaert:2015xua,Ciambelli:2019lap}. It allows us to split the coordinates into vertical and horizontal parts and help to generate the proper time dependence of the coordinate time, but then the local boost invariance gets lost. So whether it is a necessary component of the intrinsic Carrollian gravity is questionable.} instead of a (weak) Carrollian structure. As we mentioned above, a Carrollian structure should have the local boost symmetry as the gauge symmetry, but a ruled Carrollian structure requires to fix a local boost frame  and breaks the gauge symmetry. We prefer to avoid using the ruled Carrollian structure, in order to keep the gauge symmetry.  In the following sections, our result shows that even without using a ruled Carrollian structure, we can still fix the ambiguity in proper-time dependence of the coordinate time and produce well-defined geodesics. For exact neutral objects in magnetic-electric Carrollian RN black hole, we can determine the projection of their geodesics onto the `absolute space', where the projected geodesics are unique and well-defined, which might be a proper reflection of the Carrollian perspective of spacetime. When considering the motion of neutral objects in electric-electric Carrolian RN black hole, or the motion of a charged particle in the background of either type of the Carrollian RN black hole, we do not find such ambiguity, regardless of how small the charge is. We believe our work provides another viewpoint to solve the puzzle in Carrollian geodesics.

\section{Geodesics in Carrollian RN black holes}\label{sec:CarrBH}
    A Carrollian RN black hole can be viewed as the solution of electric/magnetic Carrollian gravity coupled to electric/magnetic Carrollian Maxwell electromagnetism.
    In \cite{deBoer:2023fnj}, the metrics of Carrollian RN black hole have been obtained by taking different limits of the metric of relativistic RN.   The metric of $(3+1)$-dim. relativistic RN black hole is
    \begin{equation}
        ds^2 = -\left(1 - \frac{R_S}{r} + \frac{R^2_{Q,P}}{r^2}\right)dt^2 + \left(1 - \frac{R_S}{r} + \frac{R^2_{Q,P}}{r^2}\right)^{-1}dr^2 + r^2(d\theta^2 + \sin^2\theta \, d\phi^2),
    \end{equation} where
    \begin{equation}
        R_S=\frac{2G_N M}{c^2}, \quad R^2_{Q,P}=\frac{1}{4\pi}\left(\frac{Q^2}{\epsilon_0}+P^2 \mu_0\right)\frac{G_N}{c^4}, \quad x^\mu= (t,r,\theta,\phi) .
    \end{equation}
    A relativistic RN black hole can have electric and/or magnetic charges due to the electromagnetic duality in relativistic electromagnetism. In Carrollian electromagnetic theory, this electromagnetic duality differs significantly from the relativistic case, as discussed in \cite{Chen:2024vho}. Therefore, the meaning of magnetic charges in the Carrollian theories remains unclear. Due to  this reason, we will only consider electrically charged RN black holes and their Carrollian limits in this work, which means that we always take the electric limit on the Maxwell electromagnetism. Moreover, since the null geodesic equations \eqref{eq:nullgeo} just tell us that the null  particle does not move, we will focus on the massive cases.
    
\subsection{Electric-electric Carrollian RN black hole}
    Following the same procedure as in \cite{deBoer:2023fnj}, with $E=Mc^2$, $G^{(el)}_c=\frac{G_N}{c^2}$ and $\frac{Q^2}{\epsilon_0}$ being fixed, we take the electric limit of both Einstein and Maxwell theories, which leads to the electric-electric limit of the RN black hole. We find
    \begin{equation}
    \begin{aligned}
        \bm{n} &= -\left[ \frac{2G_cE}{r}\left(1-\frac{Q^2}{8\pi \epsilon_0E }\frac{1}{r}\right)\right]^\frac{1}{2}\frac{\partial}{\partial r}, \\
        \bm{h} &= \frac{2G_cE}{r}\left( 1-\frac{Q^2}{8\pi \epsilon_0 E}\frac{1}{r} \right) dt^2+r^2d\Omega^2.
    \end{aligned}
    \end{equation}
    Without loss of generality we can assume $\dot \theta=0$ such that the motion happens in the equatorial plane $\theta=\frac{\pi}{2}$.  After introducing 
    \begin{equation}
        a\equiv 2EG_c,~~~~ b\equiv \frac{Q^2}{8 \pi \epsilon_0 E},
    \end{equation} we obtain the action of Carrollian geodesic
    \begin{equation}
        I= \frac{1}{2} \int d\tau~e\left(m^2 + e^{-2} \left( \frac{a}{r} (1-\frac{b}{r}) \dot t^2+r^2 \dot \phi^2 \right) \right).
    \end{equation} Taking the variation with respect to $\phi$ produces a constant of motion that is the angular momentum $l$ satisfying
    \begin{equation}\label{eq:EEeq-phi-2}
        \dot \phi=\frac{el}{r^2}.
    \end{equation} 
    We should keep in mind that $e$ can be chosen freely  due to the reparameterization invariance, and the common choice is to set $e=\text{constant}$ and define $\epsilon\equiv e^2m^2$. It is also easy to observe that $\frac{\partial}{\partial t}$ is a Killing vector, and by the  equation of motion of $t$,
    \begin{equation}\label{eq:EEeq-t}
        \frac{d}{d \tau} \left(\frac{1}{e}\frac{2a}{r} (1-\frac{b}{r}) \dot t \right)=0 ~~~~\Rightarrow ~~~~\dot t = \frac{e\Sigma}{\frac{a}{r}(1-\frac{b}{r})},
    \end{equation} 
    where $\Sigma$ is the conserved charge corresponding to the translation invariance along $t$. Now varying the action with respect to $r$, we get
    \begin{equation}
        2r \dot \phi^2-\frac{a}{r^2}(1-\frac{2b}{r}) \dot t^2=0.
    \end{equation}
    This equation  is actually equivalent to affine reparameterization invariance when \eqref{eq:EEeq-phi-2} and \eqref{eq:EEeq-t} hold
    \begin{equation}\label{eq:EEeq-affine}
        e^2 m^2 = \left( \frac{a}{r} (1-\frac{b}{r}) \dot t^2+r^2 \dot \phi^2 \right) =~\text{const}.
    \end{equation}
    Substituting \eqref{eq:EEeq-phi-2} and \eqref{eq:EEeq-t} into \eqref{eq:EEeq-affine}, we have
    \begin{equation}\label{eq:EEeq-r}
        \frac{2al^2}{r^3}(1-\frac{b}{r})^2=(1-\frac{2b}{r}) \Sigma^2.
    \end{equation}
    We can see that there exist two constants of motion, the angular momentum $l$ and $\Sigma$, with both of which completely determining the radial  coordinate $r$ via \eqref{eq:EEeq-r}. But as it is a polynomial equation, $r$ would only admit finite number of discrete values, and when these solutions exist we have both $t,\phi$ linearly depending on $\tau$. Considering that $\bm{n} \propto \frac{\partial}{\partial r}$ for electric-electric RN black hole, $r$ is somehow time-like, so in this context these solutions describe the objects behaving as `instantons' winding around the space with topology $\mathbb{R}_t \cross \mathbb{S}^2_{\theta,\phi}$.

\subsection{Magnetic-electric Carrollian RN black hole}
    With $E=Mc^2$, $G^{(m)}_c=\frac{G_N}{c^4}$ and $\frac{Q^2}{\epsilon_0}$ being fixed, we can take the magnetic limit of the Einstein theory and the electric limit of the Maxwell theory, which leads to the magnetic-electric limit of the RN black hole:
    \begin{equation}
    \begin{aligned}
        \bm{n}&= - \left( 1-\frac{R_S}{r}+\frac{R^2_Q}{r^2} \right)^{-\frac{1}{2}}\frac{\partial}{\partial t},\\
        \bm{h}&= \left( 1-\frac{R_S}{r}+\frac{R^2_Q}{r^2} \right)^{-1}dr^2+r^2d\Omega^2,
    \end{aligned}
    \end{equation} where $R^2_Q=\frac{1}{4\pi \epsilon_0}Q^2G^{m}_c$.

    Again without loss of generality, we constrain the motion in the equatorial plane $\theta=\frac{\pi}{2}$ and use the conservation of angular momentum and affine parameterization invariance to fix $\epsilon=e^2 m^2>0$ and reduce the equation of motion to
    \begin{equation}
        \frac{\dot r^2}{2}+V_{eff}(r)=\frac{\epsilon}{2}.
    \end{equation}  Here the effective potential is 
    \begin{equation}
        \begin{aligned}
            V_{eff}(r)&=\frac{\epsilon}{2}-\frac{1}{2}\left(\epsilon-\frac{e^2 l^2}{r^2}\right)\left(1-\frac{R_S}{r}+\frac{R^2_Q}{r^2}\right)\\
            &=\frac{R_S\epsilon}{2r}-\frac{R^2_Q\epsilon}{2r^2}+\frac{e^2l^2}{2r^2}-\frac{R_Se^2l^2}{2r^3}+\frac{R^2_Qe^2l^2}{2r^4}.
        \end{aligned}
    \end{equation}
    Remarkably, there is only one small but significant difference from the effective potential for the geodesics in the relativistic RN black hole: the first two Newtonian terms now depend linearly on the test particle's affine parameterization constant. We can recast the radial equation of motion into the form
    \begin{equation}\label{eq:ME-constraint}
        \dot r^2=(1-\frac{R_S}{r}+\frac{R^2_Q}{r^2})(\frac{1}{b_{in}^2}-\frac{1}{r^2})e^2l^2
    \end{equation} where $b_{in}=\frac{l}{m}$ determines the asymptotic radial velocity. We can see that as $\dot r^2 \geq 0$, this equation  puts a strong constraint on possible geodesics. Furthermore, as we have mentioned, the above is all the information about the geodesics that we can derive, and $t(\tau)$ is totally undetermined. But since the Carrollian manifold has a canonical bundle structure specified by $n^\mu$, we can project everything onto its base manifold, which has the meaning of absolute space, and draw the geodesics parameterized by $(r,\theta,\phi)$. The projected geodesics are unique and well-defined.

    Let us first discuss the circular orbits. For the circular orbits, we have $V_{eff}=0$ and $\frac{dV_{eff}}{dr}=0$, which lead to
    \begin{equation}\label{eq:circularorbit}
    \begin{aligned}
        (1-\frac{R_S}{r}+\frac{R^2_Q}{r^2})(\frac{1}{b_{in}^2}-\frac{1}{r^2})l^2&=0,\\
        \frac{l^2}{r^2}\left(\frac{R_S}{2b_{in}^2}-\frac{R^2_Q}{rb_{in}^2}+\frac{1}{r}-\frac{3R_S}{2r^2}+\frac{2R^2_Q}{r^3}\right)&=0.
    \end{aligned}
    \end{equation}
    By solving them we can easily see that when $\Delta=R^2_S-4R^2_Q>0$ we have $r=b_{in}=r_\pm$, which represents the two solutions of the quadratic equation $1-\frac{R_S}{r}+\frac{R^2_Q}{r^2}=0$. Thus, all possible circular orbits sit at the Carroll extremal surface (CES) \cite{Ecker:2023uwm} $r=r_\pm$ when $b_{in}=r_\pm$ respectively. Moreover, as
    \begin{equation}
        \left.\frac{d^2V_{eff}}{d^2r}\right|_{r=b_{in}=r_\pm}=\frac{2l^2}{r^6_\pm}(R^2_Q-r^2_\pm),
    \end{equation} we conclude that when $\Delta>0$, the circular orbit at $r=r_+$ is unstable, while the circular orbit at $r=r_-$ is stable. 
    
    For the extreme black hole with $\Delta=0$, $r=r_+=r_-=\frac{R_S}{2}$ is always a solution to \eqref{eq:circularorbit}, independent of $b_{in}$. As
    \begin{equation}
       \left. \frac{d^2V_{eff}}{d^2r}\right|_{r=\frac{R_S}{2}}= \frac{-4l^2}{R_S^2}\left(\frac{1}{b_{in}^2}-\frac{4}{R_S^2}\right),
    \end{equation} 
    the circular orbit at $r=\frac{R_S}{2}$ is stable when $b_{in}> \frac{R_S}{2}$, and it is unstable when $b_{in} \leq \frac{R_S}{2}$. The stability when $b_{in}=\frac{R_S}{2}$ can be checked according to the third derivative
    \begin{equation}
        \left.\frac{d^3V_{eff}}{d^3r}\right|_{r=b_{in}=\frac{R_S}{2}}=\frac{-192l^2}{R_S^5} < 0,
    \end{equation} which suggests that the circular  motion is unstable.

     In fact, when $\Delta=R^2_S-4R^2_Q\geq0$ there could also be bounded orbits other than circular orbits.   To discuss them carefully we need to do analytic continuation to understand what happens near the CES.
    Following the discussion of the Schwarzschild Carrollian wormhole in \cite{deBoer:2023fnj}, in the case of magnetic-electric Carrollian RN black hole, we can introduce a new radial coordinate $\rho$ in the regions $r\geq r_+$ and $0<r\leq r_-$ such that
    \begin{equation}\label{eq:ME outer}
        \frac{d\rho}{\rho}=\frac{dr}{\sqrt{(r-r_+)(r-r_-)}}
    \end{equation}
    Now we can find that the metric is conformally flat
    \begin{equation}
        \bm{h}=\frac{r^2(\rho)}{\rho^2}\left(d\rho^2+\rho^2 d\Omega^2\right),
    \end{equation} where $r$ is a function of $\rho$
    \begin{equation}\label{eq:MEout-r(rho)}
    \begin{aligned}
        &r=\frac{1}{2}(r_+(\alpha+1)-r_-(\alpha-1)),\\ 
        &\alpha=\frac{1}{2}(\rho+\frac{1}{\rho}).
    \end{aligned}
    \end{equation}
    It is easy to see either $\alpha\geq1$ or $\alpha\leq-1$,  up to the  sign of $\rho$. For positive $\rho$, $r$ takes value in the region $r\geq r_+$, and $r\to\infty$ corresponds to two different asymptotic flat patches, one with $\rho \to \infty$ and the other one with $\rho \to 0$. For negative $\rho$, when $-\frac{r_+-r_-}{r_++r_-}<\alpha \leq -1$ we have $0<r\leq r_-$, and the singularity at $r \to 0$ is represented by two distinct $\rho$'s.

    We can introduce a new radial coordinate $\rho$ for the region $r_-<r<r_+$ as well, 
    \begin{equation}\label{eq:ME inner}
        \frac{d\rho}{\rho}=\frac{dr}{\sqrt{(r_+-r)(r-r_-)}}
    \end{equation}
    and get
    \begin{equation}\label{eq:MEin-r(rho)}
    \begin{aligned}
        &r=\frac{1}{2}(r_+(\alpha+1)-r_-(\alpha-1)),\\ 
        &\alpha=\cos \log \rho.
    \end{aligned}
    \end{equation} Here $-1<\alpha<1$ and $\rho$ takes positive value. Due to the periodicity of cosine function, the region $r_-<r<r_+$ is analytically continued to infinite patches in $\rho$. 

    Now we start the discussion on the bounded orbits for $\Delta>0$ and mostly focus on the radial motion. The same discussion about extreme black hole with $\Delta=0$ is put off to the next part about deflection angle for convenience. 
    \begin{itemize}
    \item When $b_{in}>r_+$, according to \eqref{eq:ME-constraint} the particle motion would be constrained in $r_-\leq r\leq r_+$, and according to \eqref{eq:ME inner} $\dot \rho \neq 0$ at the CESs.  The particle travels between the two CESs, and get to a different patch in $\rho$ (see \eqref{eq:MEin-r(rho)})  each time crossing any CES.

    \item In the critical situation $b_{in}=r_+$, the bounded motion is also constrained in $r_-\leq r\leq r_+$, but in this case $\dot \rho=0$ at $r=r_+$.  We have already seen that there is an unstable circular orbit at the outer CES in this case, so after an inward perturbation the particle falls from the unstable circular orbit to the inner CES, gets to another patch and finally returns to the outer CES. However, it costs the particle infinite proper time to reach or leave the outer CES because $\dot r\sim (r-r_+)$ near $r=r_+$.  In the critical case, the particle would need to wind an infinite number of times to reach or leave the outer CES. We will give a more detailed discussion  about unbounded orbits below.

    \item When $r_-<b_{in}<r_+$, the bounded motion is constrained in $r_-\leq r \leq b_{in}$, and we have $\dot \rho \neq0$ at $r=r_-$ while $\dot \rho=0$ at $r=b_{in}$. So the particle would travel between two adjacent patches through the inner CES with the turning points at $r=b_{in}$. 

    \item When $b_{in}=r_-$, the only allowed bounded motion is the circular orbit at $r=r_-$.

    \item For $0\leq b_{in}<r_-$, the bounded motion is allowed in $b_{in}\leq r\leq r_-$, which is inside the inner CES. Similar to the $r_-<b_{in}<r_+$ case, the particle would travel between the two patches of negative $\rho$ \eqref{eq:MEout-r(rho)} and the turning points sit at $r=b_{in}$. In particular, only the particles with $b_{in}=0$ will fall into the singularity.

    \end{itemize}

    Next, we turn to the unbounded orbits and pay special attention to the deflection angle in this class of orbits if possible. The deflection angle is the measure of how much a light ray is bent as it passes near a massive object. Since in a Carrollian background the massless geodesics never get bent \eqref{eq:nullgeo}, one may consider the massive case instead. The deflection angle in the presence of a Carrollian Schwarzschild black hole has been calculated in \cite{Ciambelli:2023tzb}, and now we discuss it in the Carrollian RN black hole. A special point is that the notion of the deflection angle can be well-defined after  projection onto the absolute space, making it intrinsic to Carrollian geometry. Let us discuss $\Delta>0$ and $\Delta=0$ case by case.

    \paragraph{{\bf 1. $\Delta>0$ case.}}
    
    When $\Delta>0$, as we know
    \begin{equation}
        \frac{d\phi}{dr}=\frac{\dot\phi}{\dot r}=\pm \frac{b_{in}}{\sqrt{(r^2-b_{in}^2)(r^2-R_Sr+R^2_Q)}},
    \end{equation} by introducing $u=1/r$, we get the integral form of the change in angle  as
    \begin{equation}\label{phiinfty}
        \phi_\infty=2\int^{u_0}_0\frac{b_{in}du}{\sqrt{(1-b_{in}^2u^2)(1-R_Su+R^2_Qu^2)}}.
    \end{equation} Demanding $\frac{du}{d\phi}\mid_{u_0}=0$, we see that $u_0=\min\lbrace\frac{1}{b_{in}},\frac{1}{r_\pm}\rbrace$. The deflection angle would be \begin{equation}
        \Delta \phi\equiv \phi_\infty-\pi.
    \end{equation} Then we may classify these geodesics according to the incident parameter $b_{in}$.
    
    \begin{itemize}
        \item For the particles incident from infinity with $b_{in}>r_+$, they will be bounced  back to infinity before reaching the CES, so we are able to discuss the deflection angle. Specially, if both $R_S$ and $R^2_Q$ vanish,  we have
    \begin{equation}
        \phi_\infty=2\arcsin(b_{in}u_0) = \pi.
    \end{equation} 
    There is no deflection, as expected. In general, 
    introducing 
    \begin{equation}
      x\equiv \frac{r_-}{b_{in}},\hspace{3ex}  y\equiv \frac{r_+}{b_{in}},
    \end{equation}  we may write the above integral \eqref{phiinfty} in terms of elliptic integrals
    \begin{equation}\label{eq:deflection-ang}
        \phi_\infty=\frac{4}{\sqrt{(1-x)(1+y)}}\left(-F(\arcsin\sqrt{\frac{1-x}{2}},-\frac{2(x-y)}{(1-x)(1+y)})+K(-\frac{2(x-y)}{(1-x)(1+y)})\right).
    \end{equation} 
    For $b_{in}\gg r_+>r_-$, expanding $\phi_\infty$ near $x=y=0$ we get
    \begin{equation}
        \phi_\infty=\pi+x+y+\frac{3\pi y^2}{16}+\frac{\pi xy}{8}+\frac{3\pi x^2}{16}+\frac{5y^3}{12}+\frac{xy^2}{4}+\frac{x^2y}{4}+\frac{5x^3}{12}+\cdots,
    \end{equation} 
    By $x+y=\frac{r_++r_-}{b_{in}}=\frac{R_S}{b_{in}}$ and $ xy=\frac{r_+r_-}{b_{in}^2}=\frac{R^2_Q}{b_{in}^2}$, we see that $R_S$ starts to appear in the terms of order $\mathcal{O}(x)$, while $R^2_Q$ appears in $\mathcal{O}(x^2)$ terms, and the first mixed term $R_SR^2_Q$ appears in $\mathcal{O}(x^3)$ terms. So to order of $\mathcal{O}(x^2)$ we have
    \begin{equation}
        \phi_\infty-\pi = \frac{R_S}{b_{in}}-\frac{\pi R^2_Q}{b_{in}^2}+\frac{3\pi R^2_S}{16b_{in}^2} + \mathcal{O}(x^3).
    \end{equation} 
    The leading order $\frac{R_S}{b_{in}}$ is not corrected by the electromagnetic effects and matches the leading-order result in the case of Carrollian Schwarzschild black hole\cite{Ciambelli:2023tzb}. The next-to-leading-order correction is influenced by both $R_S$ and $R^2_Q$. Moreover, at order $\mathcal{O}(x^3)$, we find the correction is
    \begin{equation}
        \frac{5y^3}{12}+\frac{xy^2}{4}+\frac{x^2y}{4}+\frac{5x^3}{12}=\frac{5}{12}\left(\frac{R_S}{b_{in}}\right)^3-\frac{R^2_QR_S}{b_{in}^3}.
    \end{equation} 
     In the case that $b_{in}$ is slightly larger than $r_+$, the particle could wind around the black hole several times before returning to infinity, as shown in Figure \ref{geodesic-fig}. Now we may expand \eqref{eq:deflection-ang} as $y \to 1^-$ and take $k =\frac{r_-}{r_+},$
    \begin{equation}
    \begin{aligned}
        \phi_\infty&= -\sqrt{\frac{2}{1-k}}\ln(1-y) + \gamma  + \mathcal{O}(1-y),\\
        \gamma &= \sqrt{\frac{2}{1-k}}\ln \left(\frac{32(1-k)}{\left(\sqrt{1-k}+\sqrt{2}\right)^2}\right) .
    \end{aligned}
    \end{equation} 
    This means that the parameter $b_{in}$ is related to the winding number $n$ by
    \begin{equation}
        b_{in}(n) \approx \frac{r_+}{1- e^{-\sqrt{\frac{1-k}{2}}\left(2\pi n-\gamma\right)}}.
    \end{equation} 
    When $R_Q^2=0$, it recovers the same expression for the Carrollian Schwarzschild black hole \cite{Ciambelli:2023tzb} after  setting $k=0$:
    \begin{equation}
    \begin{aligned}
        b_{in}(n) &\approx \frac{R_S}{1- e^{-\frac{1}{\sqrt{2}}\left(2\pi n-\gamma\right)}},\\
        \gamma &= \sqrt{2}\ln \left(\frac{32}{\left(1+\sqrt{2}\right)^2}\right) .
    \end{aligned}
    \end{equation} 
     As the winding number increases, the impact parameter $b_{in}$ of the incident particle gets exponentially close to  the outer CES radius $r_+$. The discussion here is similar to the discussion about the photon sphere or photon ring in GR, except that now we are discussing massive particles.

    \begin{figure}[htbp]
    \centering
    \subfloat[\centering $b_{in}>r_+$]
    {
        \begin{minipage}[b]{.45\linewidth}
            \centering
            \includegraphics[scale=0.15]{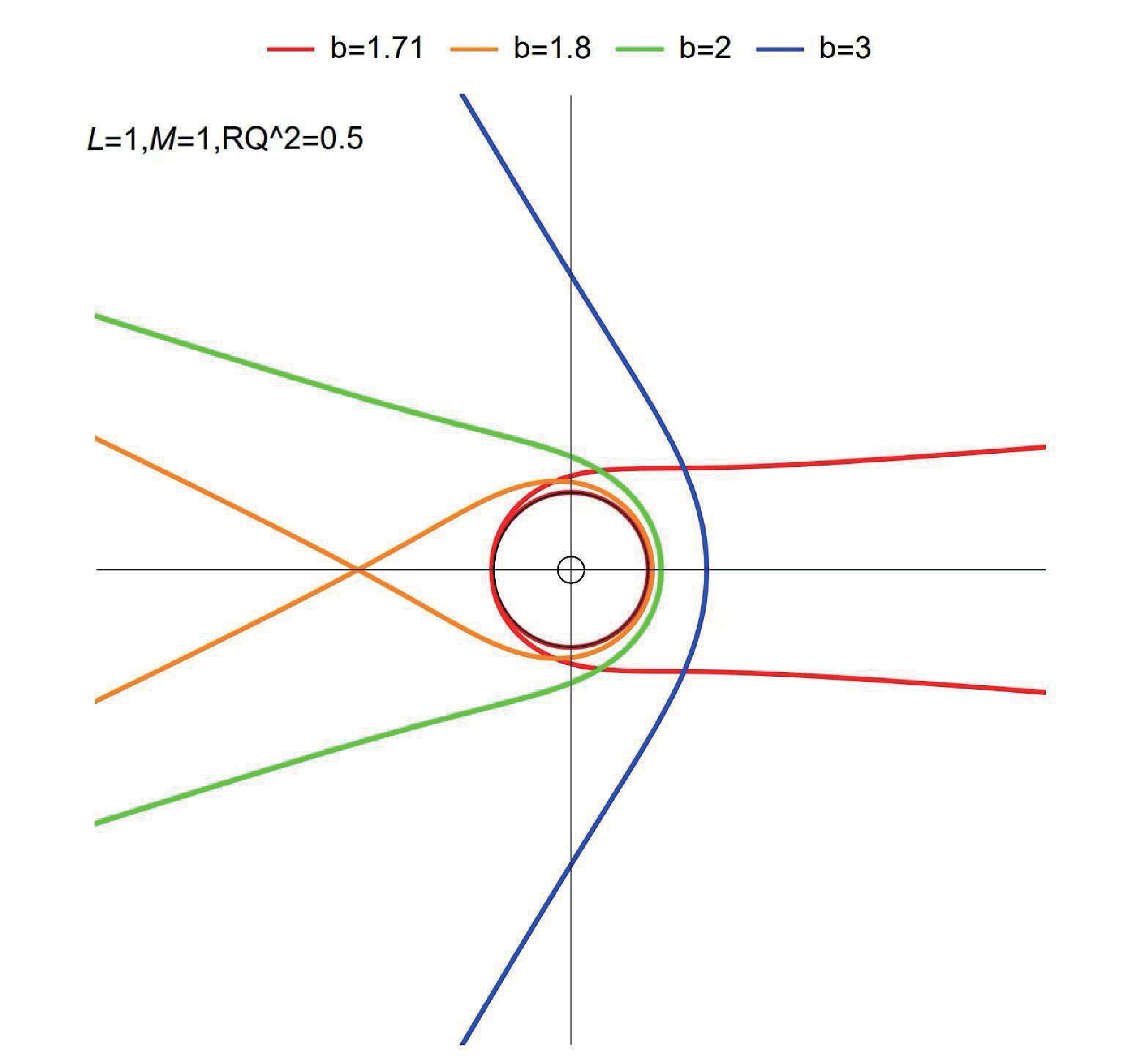}
        \end{minipage}
    }
    \subfloat[\centering $b_{in}<r_+$]
    {
     	\begin{minipage}[b]{.45\linewidth}
            \centering
            \includegraphics[scale=0.15]{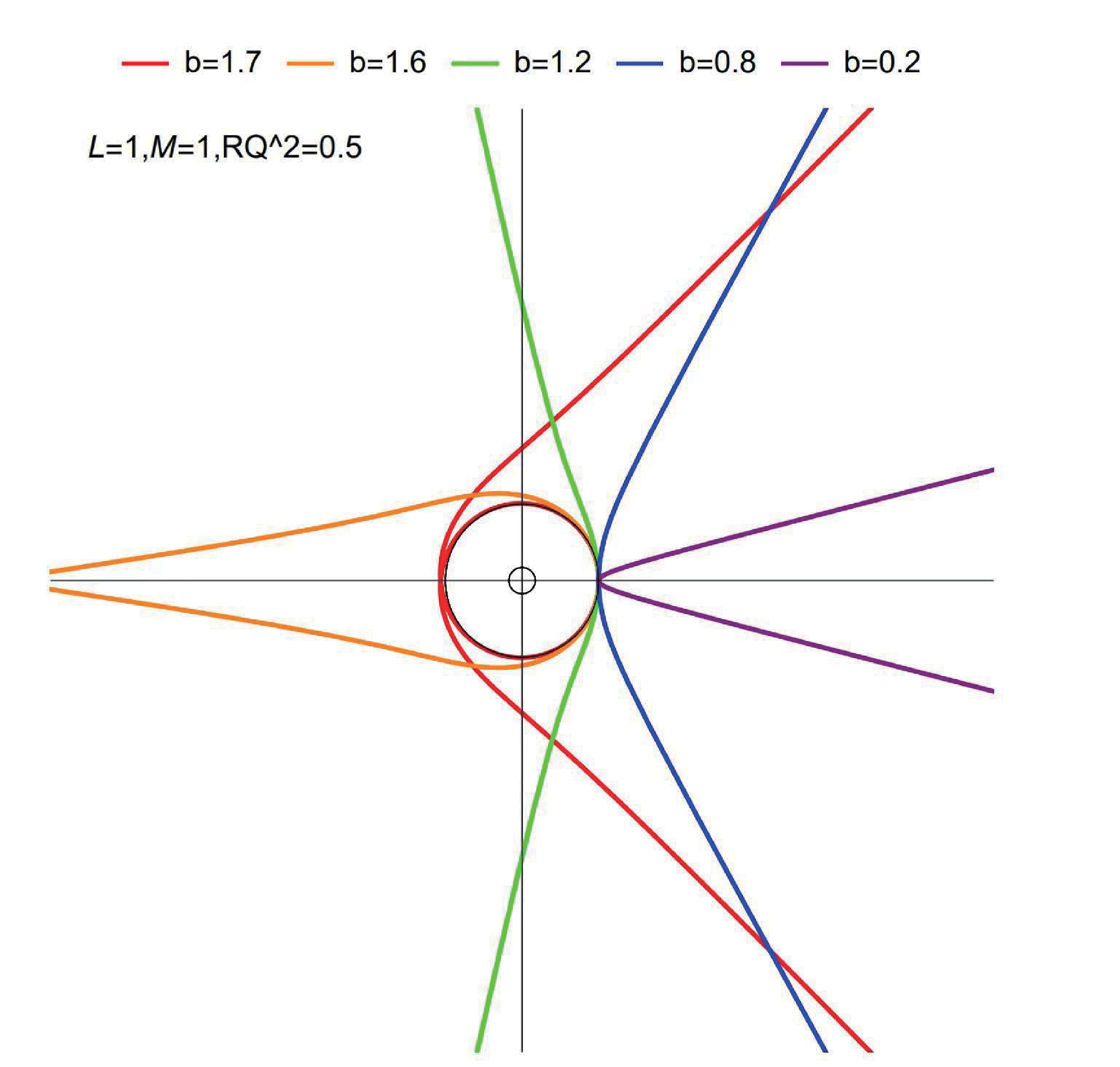}
        \end{minipage}
    }
    \caption{\centering \label{geodesic-fig} Geodesics of infalling particles from infinity for fixed $R_S,R_Q^2$ and different $b_{in}$}
    \end{figure}

    \item Next, consider the  critical situation $b_{in}=r_+$. In this case, the particle incident from infinity would wind around infinite times before reaching the unstable circular orbit at $r=r_+$. It takes infinite proper time for the particle to reach the orbit. On the other hand, as discussed above, in this case there also exists bounded motion in $r_-\leq r\leq r_+$ and unstable circular orbit at $r = r_+$. This means that after an outward perturbation the particle leaves from the unstable circular orbit to infinity, but it also costs the particle infinite proper time to leave because the winding number for $b_{in}=R_S$ is divergent. More precisely, for a particle moving on the unstable circular orbit, it may leave the orbit under a perturbation,   either to the asymptotic flat patch \eqref{eq:MEout-r(rho)} $0<\rho<1$ or into $\rho>1$ depending on the direction of the perturbation. 

    \item When $b_{in}<r_+$, the constraint \eqref{eq:ME-constraint} implies  unbounded motions  in the region $r\geq r_+ $. Near $r=r_+$, $\dot{r}\sim (r-r_+)^\frac{1}{2}$,  it just takes a particle finite proper time to reach  the outer CES. However, at $r=r_+$, $\dot \rho \neq 0$, and thus the inextensible geodesic will reach a different asymptotic flat patch from the original one. It makes no sense to  discuss the deflection angle because the particle never comes back. However, we can still count the change in angle before the particle hits the outer CES, which amounts to half of $\phi_\infty$,
    \begin{equation}
        \frac{\phi_\infty}{2}=\int^{\frac{1}{r_+}}_0\frac{b_{in}du}{\sqrt{(1-b_{in}^2u^2)(1-R_Su+R^2_Qu^2)}}.
    \end{equation} It is extremely hard to write down its explicit expression, but when $b_{in}=r_-$ it simplifies to
    \begin{equation}
    \begin{aligned}
        \frac{\phi_\infty}{2}&=\int^{\frac{1}{r_+}}_0\frac{b_{in}du}{\sqrt{(1-r_-^2u^2)(1-r_+u)(1-r_-u)}}\\
        &=\sqrt{\frac{2k}{1-k}}\arctan\sqrt{\frac{2k}{1-k}},
    \end{aligned}
    \end{equation} where $k=\frac{r_-}{r_+} \in [0,1)$ characterizes how close the black hole is to the extreme Carrollian RN black hole. It is easy to verify $\frac{d\phi_\infty}{d k}<0$, which means that the closer the black hole is to the extreme Carrollian RN black hole, the larger $\phi_\infty$ is. Actually, for the black holes with $\Delta>0$, $\phi_\infty$ is always finite when $b_{in}<r_+$, but it diverges for the extreme Carrollian RN black hole.
    
    \end{itemize}

    As we can see, as in the case of the Carrollian Schwarzschild black hole, in the nonextreme Carrollian RN black hole there exist geodesics in which a particle incidents from infinity, hits the outer CES and then makes a U-turn back to another asymptotic flat  region  in finite proper time. Now the black hole does not absorb anything, but acts as a mirror.

\paragraph{{\bf 2. $\Delta =0$ case.}}

    Now we consider the extreme Carrollian RN black hole with $r_+=r_-=\frac{R_S}{2}$. The radial equation is
    \begin{equation}
        \dot r^2=(1-\frac{R_S}{2r})^2(\frac{1}{b_{in}^2}-\frac{1}{r^2})e^2l^2\geq0
    \end{equation} 
    For an extreme RN black hole, again we introduce a new radial coordinate $\rho$ such that
    \begin{equation}
        \frac{d\rho}{\abs{\rho}}=\frac{dr}{\abs{r-\frac{R_S}{2}}},
    \end{equation}
    and we could have
    \begin{equation}
    \begin{aligned}
        r&=a\rho+\frac{R_S}{2},
        \\\bm{h}&=\frac{r^2(\rho)}{\rho^2}\left(d\rho^2+\rho^2 d\Omega^2\right).
    \end{aligned}  
    \end{equation} 
    Here $a$ is an arbitrary constant. As $-\frac{R_s}{2a}<\rho<0$ corresponds to the region $0<r<\frac{R_s}{2}$, while $\rho>\frac{R_s}{2a}$ corresponds to $r>\frac{R_s}{2}$,  there is only one asymptotic flat patch.
    
    For the deflection angle in this case, we have 
    \begin{equation}\label{eq:delta=0}
    \begin{aligned}
        \frac{d\phi}{dr}&=\pm \frac{b_{in}}{\sqrt{(r^2-b_{in}^2)(r-\frac{R_S}{2})^2}},\\
        \phi_\infty&=2\int^{u_0}_0\frac{b_{in}du}{\sqrt{(1-b_{in}^2u^2)(1-\frac{R_S}{2}u)^2}},
    \end{aligned}
    \end{equation} 
    where $u_0=\mbox{min}\lbrace\frac{1}{b_{in}},\frac{2}{R_S}\rbrace$. The explicit  expression of this integral is
    \begin{equation}
        \phi_\infty=\frac{2 \arcsin (x)+\pi }{\sqrt{1-x^2}},~~~~~
        \mbox{if}~ x=\frac{R_S}{2b_{in}} \in (0,1),
    \end{equation}
     and otherwise it diverges.

    \begin{itemize}
    \item When $b_{in} \gg \frac{R_S}{2}$, expanding near $x=0$, we get
    \begin{equation}
    \begin{aligned}
        \phi_\infty&=\pi+2x+\frac{\pi}{2}x^2+\frac{4}{3}x^3+\cdots\\
        &=\pi+\frac{R_S}{b_{in}}+\frac{R_S^2}{8b_{in}^2}+\frac{R_S^3}{6b_{in}^3}+\cdots
    \end{aligned}
    \end{equation} which is consistent with our earlier result when $\Delta>0$ by setting $R^2_S=4R^2_Q$. 
    
    \item When $b_{in}$ is slightly larger than $ \frac{R_S}{2}$, the expansion for $x\to1^-$ yields
    \begin{equation}
        \phi_\infty= \sqrt{\frac{2}{1-x}}\pi -2 +\mathcal{O}(\sqrt{1-x}).
    \end{equation} This implies that $b_{in}$ is related to the winding number $n$ as
    \begin{equation}
        b_{in}(n) \approx \frac{R_S}{2-\frac{1}{\left(n+\frac{1}{\pi}\right)^2}}.
    \end{equation} Now the the particle approaches the CES as the winding number increases in a power-law form.

    \item When $b_{in}=\frac{R_S}{2}$, the constraint equation implies $r\geq \frac{R_S}{2}$. And near $r=\frac{R_S}{2}$, $\dot r\sim (r-\frac{R_S}{2})^\frac{3}{2}$. The particle at the unstable circular orbit $r=\frac{R_S}{2}$ could only move to the infinity after perturbation. 

    \item When $b_{in}<\frac{R_S}{2}$, as near $r=\frac{R_S}{2}$, $\dot r\sim (r-\frac{R_S}{2})$, the particles incident at infinity would take infinite proper time to reach the CES at $r=\frac{R_S}{2}$ and also has a divergent winding number. To see this we can cut off the upper limit of the integral \eqref{eq:delta=0} at a distance from the CES $r_c > \frac{R_S}{2}$ then take the limit $r_c \to \frac{R_S}{2}$:
    \begin{equation}\label{eq:EM-Extre}
        \frac{\phi_{\infty,r_c}}{2} \sim \frac{1}{2\sqrt{x^2-1}} \ln(1-\frac{R_S}{2r_c})+ \text{a finite function of $x$}
    \end{equation}
    where $x=\frac{R_S}{2 b_{in}}>1$, and we can see clearly that the winding number diverges. This picture is different from that in the $\Delta >0$ black hole where the particles with small enough $b_{in}$ will be reflected by the CES in finite proper time to another patch. This confirms that there is only one asymptotic flat patch for an extreme Carrollian RN black hole.
    \begin{figure}[htbp]
    \centering
    \includegraphics[scale=0.75]{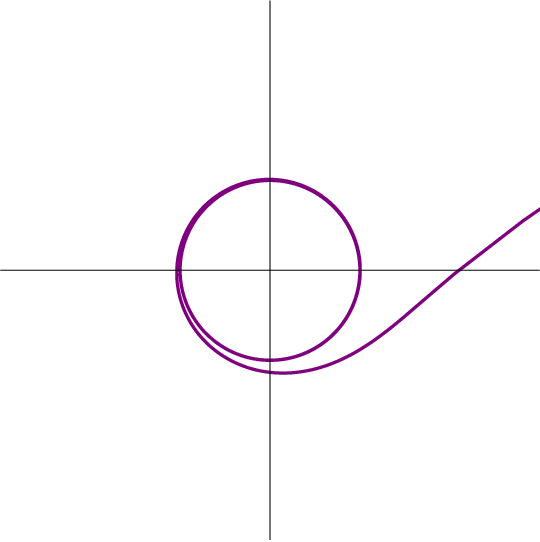}
    \caption{\centering Geodesics of infalling particles in the extreme Carrollian RN black hole with $b_{in}<\frac{R_S}{2}$}
    \end{figure}
    
    On the other hand,  the only allowed bounded orbits except the circular orbits in extreme Carrollian RN black holes is in the region $b_{in}\leq r \leq \frac{R_s}{2}$ when $b_{in}<\frac{R_s}{2}$. We notice that $r= \frac{R_S}{2}$ is the unstable circular orbit, so after an inward perturbation the particle falls to $r=b_{in}$, or leaves this orbit and moves to the infinity after an outward perturbation. But it costs the particle infinite proper time to reach or leave the CES and the winding number is divergent as \eqref{eq:EM-Extre}. This is similar to the $b_{in}=r_+$ critical situation in the non-extreme case,  but now there is one patch. 
    \end{itemize}

\subsection{Motions of Charged Particles}

    For the charged particles we can consider their minimal coupling to $A_\mu$, and construct the Carrollian invariant action as
    \begin{equation}
        I= \frac{1}{2} \int d\tau e\left( m^2 + e^{-2} h_{\mu \nu}\dot x^\mu\dot x^\nu \right) +\int d\tau q A_\mu \dot x^\mu.
    \end{equation} The geodesic equation becomes
    \begin{equation}
        h_{\sigma\lambda} \frac{d^2 x^\lambda}{d\tau^2} + \frac{1}{2}  \left( \frac{\partial h_{\sigma\mu}}{\partial x^\nu} + \frac{\partial h_{\sigma\nu}}{\partial x^\mu} - \frac{\partial h_{\mu\nu}}{\partial x^\sigma} \right) \frac{dx^\mu}{d\tau} \frac{dx^\nu}{d\tau} = \frac{q}{m} F_{\sigma\lambda} \frac{dx^\lambda}{d\tau}.
    \end{equation}
    This gives the motion of a charged particle in the Carrollian curved background.

    Let us first consider the charged particles moving in the electric-electric Carrollian RN black hole.
    In this situation we have
    \begin{equation}
        I=\frac{1}{2} \int d\tau e\left( m^2 + e^{-2} \left( \frac{a}{r} (1-\frac{b}{r}) \dot t^2+r^2 \dot \phi^2 \right) \right)-\int d\tau \frac{1}{4\pi \epsilon_0}\frac{qQ}{r}\dot t.
    \end{equation} 
    The variation with respect to $\phi$ produces a constant of motion that is the angular momentum $l$ satisfying
    \begin{equation}\label{eq:EEeq-phi}
        \dot \phi=\frac{el}{r^2},
    \end{equation} 
    where again $e$ can be chosen freely.
    Using the  equation of $t$, we now get
    \begin{equation}
        \dot t = e\frac{\Sigma+\frac{1}{4\pi \epsilon_0}\frac{qQ}{r}}{\frac{a}{r}(1-\frac{b}{r})}.
    \end{equation}
    Now by varying the action with respect to $r$, we get
    \begin{equation}
        r \dot \phi^2-\frac{a}{2r^2}(1-\frac{2b}{r}) \dot t^2+\frac{e}{4\pi \epsilon_0}\frac{qQ}{r^2}\dot t=0,
    \end{equation}
    which leads to
    \begin{equation}
        \frac{2al^2}{r^3}(1-\frac{b}{r})^2+\left(\Sigma+\frac{1}{4\pi \epsilon_0}\frac{qQ}{r}\right)\left(\frac{1}{4\pi \epsilon_0}\frac{qQ}{r}-\Sigma (1-\frac{2b}{r})\right)=0.
    \end{equation} 
    We again get the geodesics of instantons, and if we let $q=0$, it returns to \eqref{eq:EEeq-r} as it should be.

    Next we consider the charged particles moving in the mag-electric  Carrollian RN black hole.
    In this situation we have
    \begin{equation}
        I=\frac{1}{2}\int d\tau e\left( m^2 + e^{-2}\left( ( 1-\frac{R_S}{r}+\frac{R^2_Q}{r^2} )^{-1} \dot r^2+r^2 \dot \phi^2 \right)\right)-\int d\tau\frac{1}{4\pi \epsilon_0}\frac{qQ}{r}\dot t.
    \end{equation} By varying the action with respect to $t$, we get
    \begin{equation}
        \frac{d}{d\tau} (-\frac{1}{4\pi \epsilon_0}\frac{qQ}{r})=0\Rightarrow r=r_0=\text{const}.
    \end{equation} And the conservation of angular momentum tells
    \begin{equation}
        \dot \phi=\frac{el}{r_0^2}.
    \end{equation} In conclusion, the electric field restricts the particles to do circular motion at a constant radial  distance.  In this case, by varying with respect to $r$, and using $\dot r=0$, we can derive the equation for $t$, 
    \begin{equation}
        \frac{r_0 \dot \phi ^2}{e}+\frac{1}{4\pi \epsilon_0}\frac{qQ}{r_0^2}\dot t=0, 
    \end{equation}
    which is equivalent to
    \begin{equation}
        \dot t=-\frac{4\pi \epsilon_0}{qQ}\frac{el^2}{r_0}.
    \end{equation} This implies that the angular velocity of the particle in the coordinate time is given by
    \begin{equation}
        \frac{d \phi}{d t} =-\frac{qQ}{4\pi \epsilon_0}\frac{1}{lr_0} .
    \end{equation}
    As we can see, the presence of charge significantly alters the behavior of the particle's geodesic. In this case, the only allowed geodesic motion will be circular motion at any fixed radius.

\section{Conclusion}\label{Conclusion}

    In this paper, we analyzed the geodesics in  Carrollian RN black holes, focusing on both neutral and charged particles. By using only the weak Carrollian structure, we derived the geodesic equations, and considered the interaction of charged particles with the electromagnetic field.  We then explored both the electric-electric and magnetic-electric limits of the RN black hole,  and studied the geodesics in them.  As the null geodesic is trivial, we focused on the geodesics of massive particle. We classified all the possible geodesics, including the bounded and unbounded orbits. For the electric-electric Carrollian RN black hole, there exist only discrete circular orbits. On the contrary, for the magnetic-electric Carrollian RN black hole, there are both bounded and unbounded orbits. We introduced appropriate coordinates to do analytical continuation to probe its spacetime structure, which present a few remarkable properties:
    \begin{itemize}
        \item Remarkably, there is a sharp difference between the global structures of the magnetic-electric Carrollian RN black holes and the ones of usual relativistic RN black holes. For the usual RN black holes, there are infinite asymptotic flat regions. Under the ultrarelativistic limit, these asymptotic regions get squeezed: for the extreme Carrollian RN black hole, there is only one asymptotic flat region, while there are two asymptotic flat regions for the nonextreme ones. 
        \item There are circular orbits sitting at the Carroll extremal surfaces (CES). The outer circular orbit for the particles with $b_{in}=r_+$ is always unstable. Under a small perturbation, the particle on the orbit could wind around for many times  before reaching to infinity. This is similar to the photon sphere in ordinary Schwarzschild black hole, even though we are discussing `massive' particle. 
        \item For a particle incident from infinity in a nonextreme black hole, it may be bounced back or enter another asymptotic flat region, depending on the impact parameter. In particular, with a small impact parameter, it reaches CES in finite proper time,  instead of passing through the CES, it seems to be reflected but to another asymptotic flat region. In this sense, the CES acts as a perfect mirror\cite{Ciambelli:2023xqk}. 
        \item For a particle incident from infinity in an extreme black hole, it will be bounced back if the impact parameter is greater than the CES radius. But with a  impact parameter $b_{in}< \frac{R_S}{2}$, the particle will wind around the black hole infinite times and be captured by the extreme black hole near its CES instead of being reflected. Now the `photon sphere' of the extreme black hole is filled by the particles with different impact parameters ranging in $b_{in} \leq \frac{R_S}{2}$. Under a perturbation, the  particles on the unstable CES  can move outward or inward, depending on the perturbation direction, similar to the $b_{in}=r_+$ critical situation in the nonextreme black hole. This reflects  the fact that the spacetime structure of the extreme black hole is totally different from the nonextreme one.
        \item For the nonextreme black hole, in the region $r_-<r<r_+$ between two CESs, the degenerate metric $h_{\mu\nu}$ is no longer positive semi-definite, which makes it difficult to define the concept of time, because in this case we have both a null direction and a direction with negative signature. 
        \item For the nonextreme black hole,  the two CESs divide the whole spacetime into three geodesically complete regions. The particles can get to another patch of the same region, but cannot enter another region within finite proper time. But for the extreme black hole, there is only one CES, and the spacetime is divided into two geodesically complete regions and has an unique asymptotic flat infinity.
    \end{itemize} 

    We further studied the charged particle in Carrollian RN black holes. In the electric-electric Carrollian RN black hole, there is only geodesics of instantons, similar to the neutral particle. However, in the magnetic-electric Carrollian black hole, the charged particle presents very different behavior: the only allowed motion is the circular motion.

\section*{Acknowledgments}
    We are grateful to Zezhou Hu for valuable discussions. This research is supported by {NSFC Grant  No. 11735001, 12275004.}\par
    
    \vspace{2cm}

\appendix
\renewcommand{\appendixname}{Appendix~\Alph{section}}

\bibliographystyle{JHEP}
\bibliography{refs.bib}

\providecommand{\href}[2]{#2}\begingroup\raggedright\begin{thebibliography}{10}

\bibitem{Cartan:1924yea}
E.~Cartan, \emph{{Sur les vari\'et\'es \`a connexion affine et la th\'eorie de la relativit\'e g\'en\'eralis\'ee. (premi\`ere partie) (Suite).}}, {\emph{Annales Sci. Ecole Norm. Sup.} {\bfseries 41} (1924) 1}.

\bibitem{Andringa:2010it}
R.~Andringa, E.~Bergshoeff, S.~Panda and M.~de~Roo, \emph{{Newtonian Gravity and the Bargmann Algebra}}, \href{https://doi.org/10.1088/0264-9381/28/10/105011}{\emph{Class. Quant. Grav.} {\bfseries 28} (2011) 105011} [\href{https://arxiv.org/abs/1011.1145}{{\ttfamily 1011.1145}}].

\bibitem{Bennett:2021dbg}
J.~Bennett, \emph{{A pedagogical review of gravity as a gauge theory}},  \href{https://arxiv.org/abs/2104.02627}{{\ttfamily 2104.02627}}.

\bibitem{Hartong:2022lsy}
J.~Hartong, N.A.~Obers and G.~Oling, \emph{{Review on Non-Relativistic Gravity}}, \href{https://doi.org/10.3389/fphy.2023.1116888}{\emph{Front. in Phys.} {\bfseries 11} (2023) 1116888} [\href{https://arxiv.org/abs/2212.11309}{{\ttfamily 2212.11309}}].

\bibitem{Dautcourt:1997hb}
G.~Dautcourt, \emph{{On the ultrarelativistic limit of general relativity}}, {\emph{Acta Phys. Polon. B} {\bfseries 29} (1998) 1047} [\href{https://arxiv.org/abs/gr-qc/9801093}{{\ttfamily gr-qc/9801093}}].

\bibitem{Bekaert:2015xua}
X.~Bekaert and K.~Morand, \emph{{Connections and dynamical trajectories in generalised Newton-Cartan gravity II. An ambient perspective}}, \href{https://doi.org/10.1063/1.5030328}{\emph{J. Math. Phys.} {\bfseries 59} (2018) 072503} [\href{https://arxiv.org/abs/1505.03739}{{\ttfamily 1505.03739}}].

\bibitem{Hartong:2015xda}
J.~Hartong, \emph{{Gauging the Carroll Algebra and Ultra-Relativistic Gravity}}, \href{https://doi.org/10.1007/JHEP08(2015)069}{\emph{JHEP} {\bfseries 08} (2015) 069} [\href{https://arxiv.org/abs/1505.05011}{{\ttfamily 1505.05011}}].

\bibitem{Bergshoeff:2017btm}
E.~Bergshoeff, J.~Gomis, B.~Rollier, J.~Rosseel and T.~ter Veldhuis, \emph{{Carroll versus Galilei Gravity}}, \href{https://doi.org/10.1007/JHEP03(2017)165}{\emph{JHEP} {\bfseries 03} (2017) 165} [\href{https://arxiv.org/abs/1701.06156}{{\ttfamily 1701.06156}}].

\bibitem{Hansen:2021fxi}
D.~Hansen, N.A.~Obers, G.~Oling and B.T.~S\o{}gaard, \emph{{Carroll Expansion of General Relativity}}, \href{https://doi.org/10.21468/SciPostPhys.13.3.055}{\emph{SciPost Phys.} {\bfseries 13} (2022) 055} [\href{https://arxiv.org/abs/2112.12684}{{\ttfamily 2112.12684}}].

\bibitem{Henneaux:2021yzg}
M.~Henneaux and P.~Salgado-Rebolledo, \emph{{Carroll contractions of Lorentz-invariant theories}}, \href{https://doi.org/10.1007/JHEP11(2021)180}{\emph{JHEP} {\bfseries 11} (2021) 180} [\href{https://arxiv.org/abs/2109.06708}{{\ttfamily 2109.06708}}].

\bibitem{Figueroa-OFarrill:2022mcy}
J.~Figueroa-O'Farrill, E.~Have, S.~Prohazka and J.~Salzer, \emph{{The gauging procedure and carrollian gravity}}, \href{https://doi.org/10.1007/JHEP09(2022)243}{\emph{JHEP} {\bfseries 09} (2022) 243} [\href{https://arxiv.org/abs/2206.14178}{{\ttfamily 2206.14178}}].

\bibitem{Bergshoeff:2023rkk}
E.~Bergshoeff, J.~Figueroa-O'Farrill, K.~van Helden, J.~Rosseel, I.~Rotko and T.~ter Veldhuis, \emph{{$p$-brane Galilean and Carrollian Geometries and Gravities}},  \href{https://arxiv.org/abs/2308.12852}{{\ttfamily 2308.12852}}.

\bibitem{Ecker:2023uwm}
F.~Ecker, D.~Grumiller, J.~Hartong, A.~P\'erez, S.~Prohazka and R.~Troncoso, \emph{{Carroll black holes}}, \href{https://doi.org/10.21468/SciPostPhys.15.6.245}{\emph{SciPost Phys.} {\bfseries 15} (2023) 245} [\href{https://arxiv.org/abs/2308.10947}{{\ttfamily 2308.10947}}].

\bibitem{deBoer:2023fnj}
J.~de~Boer, J.~Hartong, N.A.~Obers, W.~Sybesma and S.~Vandoren, \emph{{Carroll stories}}, \href{https://doi.org/10.1007/JHEP09(2023)148}{\emph{JHEP} {\bfseries 09} (2023) 148} [\href{https://arxiv.org/abs/2307.06827}{{\ttfamily 2307.06827}}].

\bibitem{Tadros:2023teq}
P.~Tadros and I.~Kol\'a\v{r}, \emph{{Carrollian limit of quadratic gravity}},  \href{https://arxiv.org/abs/2307.13760}{{\ttfamily 2307.13760}}.

\bibitem{Tadros:2024qlo}
P.~Tadros and I.~Kol\'a\v{r}, \emph{{Uniqueness of Galilean and Carrollian limits of gravitational theories and application to higher derivative gravity}}, \href{https://doi.org/10.1103/PhysRevD.109.084019}{\emph{Phys. Rev. D} {\bfseries 109} (2024) 084019} [\href{https://arxiv.org/abs/2401.00967}{{\ttfamily 2401.00967}}].

\bibitem{Levy-Leblond:1965}
J.~Levy-Leblond, \emph{{Une nouvelle limite non-relativiste du groupe de poincaré}}, {\emph{Annales de l’IHP Physique théorique} {\bfseries 3} (1965) 1}.

\bibitem{SenGupta:1966qer}
N.D.~Sen~Gupta, \emph{{On an analogue of the Galilei group}}, \href{https://doi.org/10.1007/BF02740871}{\emph{Nuovo Cim. A} {\bfseries 44} (1966) 512}.

\bibitem{Souriau:1973}
J.~Souriau, \emph{{Ondes et radiations gravitationnelles}}, {\emph{Colloques Internationaux du CNRS} {\bfseries 220} (1973) 243}.

\bibitem{Duval:2017els}
C.~Duval, G.W.~Gibbons, P.A.~Horvathy and P.M.~Zhang, \emph{{Carroll symmetry of plane gravitational waves}}, \href{https://doi.org/10.1088/1361-6382/aa7f62}{\emph{Class. Quant. Grav.} {\bfseries 34} (2017) 175003} [\href{https://arxiv.org/abs/1702.08284}{{\ttfamily 1702.08284}}].

\bibitem{Penna:2018gfx}
R.F.~Penna, \emph{{Near-horizon Carroll symmetry and black hole Love numbers}},  \href{https://arxiv.org/abs/1812.05643}{{\ttfamily 1812.05643}}.

\bibitem{Donnay:2019jiz}
L.~Donnay and C.~Marteau, \emph{{Carrollian Physics at the Black Hole Horizon}}, \href{https://doi.org/10.1088/1361-6382/ab2fd5}{\emph{Class. Quant. Grav.} {\bfseries 36} (2019) 165002} [\href{https://arxiv.org/abs/1903.09654}{{\ttfamily 1903.09654}}].

\bibitem{Freidel:2022vjq}
L.~Freidel and P.~Jai-akson, \emph{{Carrollian hydrodynamics and symplectic structure on stretched horizons}},  \href{https://arxiv.org/abs/2211.06415}{{\ttfamily 2211.06415}}.

\bibitem{Redondo-Yuste:2022czg}
J.~Redondo-Yuste and L.~Lehner, \emph{{Non-linear black hole dynamics and Carrollian fluids}},  \href{https://arxiv.org/abs/2212.06175}{{\ttfamily 2212.06175}}.

\bibitem{deBoer:2021jej}
J.~de~Boer, J.~Hartong, N.A.~Obers, W.~Sybesma and S.~Vandoren, \emph{{Carroll Symmetry, Dark Energy and Inflation}}, \href{https://doi.org/10.3389/fphy.2022.810405}{\emph{Front. in Phys.} {\bfseries 10} (2022) 810405} [\href{https://arxiv.org/abs/2110.02319}{{\ttfamily 2110.02319}}].

\bibitem{Casalbuoni:2021fel}
R.~Casalbuoni, J.~Gomis and D.~Hidalgo, \emph{{World-Line Description of Fractons}},  \href{https://arxiv.org/abs/2107.09010}{{\ttfamily 2107.09010}}.

\bibitem{Pena-Benitez:2021ipo}
F.~Pe\~na Benitez, \emph{{Fractons, Symmetric Gauge Fields and Geometry}},  \href{https://arxiv.org/abs/2107.13884}{{\ttfamily 2107.13884}}.

\bibitem{Bidussi:2021nmp}
L.~Bidussi, J.~Hartong, E.~Have, J.~Musaeus and S.~Prohazka, \emph{{Fractons, dipole symmetries and curved spacetime}},  \href{https://arxiv.org/abs/2111.03668}{{\ttfamily 2111.03668}}.

\bibitem{Jain:2021ibh}
A.~Jain and K.~Jensen, \emph{{Fractons in curved space}},  \href{https://arxiv.org/abs/2111.03973}{{\ttfamily 2111.03973}}.

\bibitem{Figueroa-OFarrill:2023vbj}
J.~Figueroa-O'Farrill, A.~P\'erez and S.~Prohazka, \emph{{Carroll/fracton particles and their correspondence}}, \href{https://doi.org/10.1007/JHEP06(2023)207}{\emph{JHEP} {\bfseries 06} (2023) 207} [\href{https://arxiv.org/abs/2305.06730}{{\ttfamily 2305.06730}}].

\bibitem{Figueroa-OFarrill:2023qty}
J.~Figueroa-O'Farrill, A.~P\'erez and S.~Prohazka, \emph{{Quantum Carroll/fracton particles}},  \href{https://arxiv.org/abs/2307.05674}{{\ttfamily 2307.05674}}.

\bibitem{Armas:2023dcz}
J.~Armas and E.~Have, \emph{{Carrollian fluids and spontaneous breaking of boost symmetry}},  \href{https://arxiv.org/abs/2308.10594}{{\ttfamily 2308.10594}}.

\bibitem{Donnay:2022aba}
L.~Donnay, A.~Fiorucci, Y.~Herfray and R.~Ruzziconi, \emph{{Carrollian Perspective on Celestial Holography}}, \href{https://doi.org/10.1103/PhysRevLett.129.071602}{\emph{Phys. Rev. Lett.} {\bfseries 129} (2022) 071602} [\href{https://arxiv.org/abs/2202.04702}{{\ttfamily 2202.04702}}].

\bibitem{Donnay:2022wvx}
L.~Donnay, A.~Fiorucci, Y.~Herfray and R.~Ruzziconi, \emph{{Bridging Carrollian and celestial holography}}, \href{https://doi.org/10.1103/PhysRevD.107.126027}{\emph{Phys. Rev. D} {\bfseries 107} (2023) 126027} [\href{https://arxiv.org/abs/2212.12553}{{\ttfamily 2212.12553}}].

\bibitem{Bagchi:2022emh}
A.~Bagchi, S.~Banerjee, R.~Basu and S.~Dutta, \emph{{Scattering Amplitudes: Celestial and Carrollian}}, \href{https://doi.org/10.1103/PhysRevLett.128.241601}{\emph{Phys. Rev. Lett.} {\bfseries 128} (2022) 241601} [\href{https://arxiv.org/abs/2202.08438}{{\ttfamily 2202.08438}}].

\bibitem{Chen:2023naw}
B.~Chen and Z.~Hu, \emph{{Bulk reconstruction in flat holography}}, \href{https://doi.org/10.1007/JHEP03(2024)064}{\emph{JHEP} {\bfseries 03} (2024) 064} [\href{https://arxiv.org/abs/2312.13574}{{\ttfamily 2312.13574}}].

\bibitem{Bagchi:2020fpr}
A.~Bagchi, A.~Banerjee, S.~Chakrabortty, S.~Dutta and P.~Parekh, \emph{{A tale of three \textemdash{} tensionless strings and vacuum structure}}, \href{https://doi.org/10.1007/JHEP04(2020)061}{\emph{JHEP} {\bfseries 04} (2020) 061} [\href{https://arxiv.org/abs/2001.00354}{{\ttfamily 2001.00354}}].

\bibitem{Bagchi:2021rfw}
A.~Bagchi, M.~Mandlik and P.~Sharma, \emph{{Tensionless tales: vacua and critical dimensions}}, \href{https://doi.org/10.1007/JHEP08(2021)054}{\emph{JHEP} {\bfseries 08} (2021) 054} [\href{https://arxiv.org/abs/2105.09682}{{\ttfamily 2105.09682}}].

\bibitem{Chen:2023esw}
B.~Chen, Z.~Hu, Z.-f.~Yu and Y.-f.~Zheng, \emph{{Path-integral quantization of tensionless (super) string}}, \href{https://doi.org/10.1007/JHEP08(2023)133}{\emph{JHEP} {\bfseries 08} (2023) 133} [\href{https://arxiv.org/abs/2302.05975}{{\ttfamily 2302.05975}}].

\bibitem{Gomis:2023eav}
J.~Gomis and Z.~Yan, \emph{{Worldsheet formalism for decoupling limits in string theory}}, \href{https://doi.org/10.1007/JHEP07(2024)102}{\emph{JHEP} {\bfseries 07} (2024) 102} [\href{https://arxiv.org/abs/2311.10565}{{\ttfamily 2311.10565}}].

\bibitem{Basu:2018dub}
R.~Basu and U.N.~Chowdhury, \emph{{Dynamical structure of Carrollian Electrodynamics}}, \href{https://doi.org/10.1007/JHEP04(2018)111}{\emph{JHEP} {\bfseries 04} (2018) 111} [\href{https://arxiv.org/abs/1802.09366}{{\ttfamily 1802.09366}}].

\bibitem{Barducci:2018thr}
A.~Barducci, R.~Casalbuoni and J.~Gomis, \emph{{Vector SUSY models with Carroll or Galilei invariance}}, \href{https://doi.org/10.1103/PhysRevD.99.045016}{\emph{Phys. Rev. D} {\bfseries 99} (2019) 045016} [\href{https://arxiv.org/abs/1811.12672}{{\ttfamily 1811.12672}}].

\bibitem{Bagchi:2019clu}
A.~Bagchi, R.~Basu, A.~Mehra and P.~Nandi, \emph{{Field Theories on Null Manifolds}}, \href{https://doi.org/10.1007/JHEP02(2020)141}{\emph{JHEP} {\bfseries 02} (2020) 141} [\href{https://arxiv.org/abs/1912.09388}{{\ttfamily 1912.09388}}].

\bibitem{Bagchi:2019xfx}
A.~Bagchi, A.~Mehra and P.~Nandi, \emph{{Field Theories with Conformal Carrollian Symmetry}}, \href{https://doi.org/10.1007/JHEP05(2019)108}{\emph{JHEP} {\bfseries 05} (2019) 108} [\href{https://arxiv.org/abs/1901.10147}{{\ttfamily 1901.10147}}].

\bibitem{Chen:2020vvn}
B.~Chen, P.-X.~Hao, R.~Liu and Z.-F.~Yu, \emph{{On Galilean conformal bootstrap}}, \href{https://doi.org/10.1007/JHEP06(2021)112}{\emph{JHEP} {\bfseries 06} (2021) 112} [\href{https://arxiv.org/abs/2011.11092}{{\ttfamily 2011.11092}}].

\bibitem{Banerjee:2020qjj}
K.~Banerjee, R.~Basu, A.~Mehra, A.~Mohan and A.~Sharma, \emph{{Interacting Conformal Carrollian Theories: Cues from Electrodynamics}}, \href{https://doi.org/10.1103/PhysRevD.103.105001}{\emph{Phys. Rev. D} {\bfseries 103} (2021) 105001} [\href{https://arxiv.org/abs/2008.02829}{{\ttfamily 2008.02829}}].

\bibitem{Chen:2021xkw}
B.~Chen, R.~Liu and Y.-f.~Zheng, \emph{{On Higher-dimensional Carrollian and Galilean Conformal Field Theories}}, \href{https://doi.org/10.21468/SciPostPhys.14.5.088}{\emph{SciPost Phys.} {\bfseries 14} (2023) 088} [\href{https://arxiv.org/abs/2112.10514}{{\ttfamily 2112.10514}}].

\bibitem{Campoleoni:2021blr}
A.~Campoleoni and S.~Pekar, \emph{{Carrollian and Galilean conformal higher-spin algebras in any dimensions}},  \href{https://arxiv.org/abs/2110.07794}{{\ttfamily 2110.07794}}.

\bibitem{Hao:2021urq}
P.-x.~Hao, W.~Song, X.~Xie and Y.~Zhong, \emph{{BMS-invariant free scalar model}}, \href{https://doi.org/10.1103/PhysRevD.105.125005}{\emph{Phys. Rev. D} {\bfseries 105} (2022) 125005} [\href{https://arxiv.org/abs/2111.04701}{{\ttfamily 2111.04701}}].

\bibitem{Chen:2022jhx}
B.~Chen, P.-x.~Hao, R.~Liu and Z.-f.~Yu, \emph{{On Galilean Conformal Bootstrap II: $\xi=0$ sector}},  \href{https://arxiv.org/abs/2207.01474}{{\ttfamily 2207.01474}}.

\bibitem{Chen:2022cpx}
B.~Chen and R.~Liu, \emph{{The Shadow Formalism of Galilean CFT$_2$}},  \href{https://arxiv.org/abs/2203.10490}{{\ttfamily 2203.10490}}.

\bibitem{Hao:2022xhq}
P.-X.~Hao, W.~Song, Z.~Xiao and X.~Xie, \emph{{BMS-invariant free fermion models}}, \href{https://doi.org/10.1103/PhysRevD.109.025002}{\emph{Phys. Rev. D} {\bfseries 109} (2024) 025002} [\href{https://arxiv.org/abs/2211.06927}{{\ttfamily 2211.06927}}].

\bibitem{Yu:2022bcp}
Z.-f.~Yu and B.~Chen, \emph{{Free field realization of the BMS Ising model}}, \href{https://doi.org/10.1007/JHEP08(2023)116}{\emph{JHEP} {\bfseries 08} (2023) 116} [\href{https://arxiv.org/abs/2211.06926}{{\ttfamily 2211.06926}}].

\bibitem{Banerjee:2022ocj}
A.~Banerjee, S.~Dutta and S.~Mondal, \emph{{Carroll fermions in two dimensions}}, \href{https://doi.org/10.1103/PhysRevD.107.125020}{\emph{Phys. Rev. D} {\bfseries 107} (2023) 125020} [\href{https://arxiv.org/abs/2211.11639}{{\ttfamily 2211.11639}}].

\bibitem{Bagchi:2022eui}
A.~Bagchi, A.~Banerjee, R.~Basu, M.~Islam and S.~Mondal, \emph{{Magic fermions: Carroll and flat bands}}, \href{https://doi.org/10.1007/JHEP03(2023)227}{\emph{JHEP} {\bfseries 03} (2023) 227} [\href{https://arxiv.org/abs/2211.11640}{{\ttfamily 2211.11640}}].

\bibitem{Chen:2023pqf}
B.~Chen, R.~Liu, H.~Sun and Y.-f.~Zheng, \emph{{Constructing Carrollian field theories from null reduction}}, \href{https://doi.org/10.1007/JHEP11(2023)170}{\emph{JHEP} {\bfseries 11} (2023) 170} [\href{https://arxiv.org/abs/2301.06011}{{\ttfamily 2301.06011}}].

\bibitem{Islam:2023rnc}
M.~Islam, \emph{{Carrollian Yang-Mills theory}}, \href{https://doi.org/10.1007/JHEP05(2023)238}{\emph{JHEP} {\bfseries 05} (2023) 238} [\href{https://arxiv.org/abs/2301.00953}{{\ttfamily 2301.00953}}].

\bibitem{Ciambelli:2023xqk}
L.~Ciambelli, \emph{{Dynamics of Carrollian scalar fields}}, \href{https://doi.org/10.1088/1361-6382/ad5bb5}{\emph{Class. Quant. Grav.} {\bfseries 41} (2024) 165011} [\href{https://arxiv.org/abs/2311.04113}{{\ttfamily 2311.04113}}].

\bibitem{Islam:2023iju}
M.~Islam, \emph{{BRST symmetry of non-Lorentzian Yang-Mills theory}}, \href{https://doi.org/10.1016/j.physletb.2023.138438}{\emph{Phys. Lett. B} {\bfseries 849} (2024) 138438} [\href{https://arxiv.org/abs/2306.04241}{{\ttfamily 2306.04241}}].

\bibitem{Chen:2024voz}
B.~Chen, H.~Sun and Y.-f.~Zheng, \emph{{Quantization of Carrollian conformal scalar theories}},  \href{https://arxiv.org/abs/2406.17451}{{\ttfamily 2406.17451}}.

\bibitem{Cotler:2024xhb}
J.~Cotler, K.~Jensen, S.~Prohazka, A.~Raz, M.~Riegler and J.~Salzer, \emph{{Quantizing Carrollian field theories}}, \href{https://doi.org/10.1007/JHEP10(2024)049}{\emph{JHEP} {\bfseries 10} (2024) 049} [\href{https://arxiv.org/abs/2407.11971}{{\ttfamily 2407.11971}}].

\bibitem{Banerjee:2024hvb}
S.~Banerjee, R.~Basu and S.~Atul~Bhatkar, \emph{{Light transformation: A Celestial and Carrollian perspective}},  \href{https://arxiv.org/abs/2407.08379}{{\ttfamily 2407.08379}}.

\bibitem{Aggarwal:2024yxy}
A.~Aggarwal, F.~Ecker, D.~Grumiller and D.~Vassilevich, \emph{{Carroll-Hawking effect}}, \href{https://doi.org/10.1103/PhysRevD.110.L041506}{\emph{Phys. Rev. D} {\bfseries 110} (2024) L041506} [\href{https://arxiv.org/abs/2403.00073}{{\ttfamily 2403.00073}}].

\bibitem{Bagchi:2024unl}
A.~Bagchi, A.~Banerjee, S.~Mondal, D.~Mukherjee and H.~Muraki, \emph{{Beyond Wilson? Carroll from current deformations}}, \href{https://doi.org/10.1007/JHEP06(2024)215}{\emph{JHEP} {\bfseries 06} (2024) 215} [\href{https://arxiv.org/abs/2401.16482}{{\ttfamily 2401.16482}}].

\bibitem{He:2024yzx}
S.~He and X.-C.~Mao, \emph{{Irrelevant and marginal deformed BMS field theories}}, \href{https://doi.org/10.1007/JHEP04(2024)138}{\emph{JHEP} {\bfseries 04} (2024) 138} [\href{https://arxiv.org/abs/2401.09991}{{\ttfamily 2401.09991}}].

\bibitem{Bergshoeff:2024ytq}
E.A.~Bergshoeff, A.~Campoleoni, A.~Fontanella, L.~Mele and J.~Rosseel, \emph{{Carroll Fermions}}, \href{https://doi.org/10.22323/1.463.0235}{\emph{PoS} {\bfseries CORFU2023} (2024) 235}.

\bibitem{Kasikci:2023zdn}
O.~Kasikci, M.~Ozkan, Y.~Pang and U.~Zorba, \emph{{Carrollian supersymmetry and SYK-like models}}, \href{https://doi.org/10.1103/PhysRevD.110.L021702}{\emph{Phys. Rev. D} {\bfseries 110} (2024) L021702} [\href{https://arxiv.org/abs/2311.00039}{{\ttfamily 2311.00039}}].

\bibitem{Nguyen:2023miw}
K.~Nguyen, \emph{{Carrollian conformal correlators and massless scattering amplitudes}}, \href{https://doi.org/10.1007/JHEP01(2024)076}{\emph{JHEP} {\bfseries 01} (2024) 076} [\href{https://arxiv.org/abs/2311.09869}{{\ttfamily 2311.09869}}].

\bibitem{Chen:2024vho}
B.~Chen, J.~Hou and H.~Sun, \emph{{On self-dual Carrollian conformal nonlinear electrodynamics}}, \href{https://doi.org/10.1007/JHEP08(2024)160}{\emph{JHEP} {\bfseries 08} (2024) 160} [\href{https://arxiv.org/abs/2405.04105}{{\ttfamily 2405.04105}}].

\bibitem{OConnor:2024rku}
J.A.~O'Connor and S.~Pekar, \emph{{A note on non-Lorentzian duality symmetries}},  \href{https://arxiv.org/abs/2409.12279}{{\ttfamily 2409.12279}}.

\bibitem{Mehra:2024zqv}
A.~Mehra, H.~Rathi and D.~Roychowdhury, \emph{{Carrollian Born-Infeld Electrodynamics}},  \href{https://arxiv.org/abs/2401.06958}{{\ttfamily 2401.06958}}.

\bibitem{GUPTA:2024tcd}
N.~GUPTA, \emph{{Aspects of Chiral Symmetries in Holography}}, Ph.D. thesis, HBNI, Mumbai, Chennai, 6, 2024.

\bibitem{Correa:2024qej}
F.~Correa, A.~Hern\'andez and J.~Oliva, \emph{{The Carrollian limit of ModMax electrodynamics}},  \href{https://arxiv.org/abs/2409.18095}{{\ttfamily 2409.18095}}.

\bibitem{Ruzziconi:2024zkr}
R.~Ruzziconi, S.~Stieberger, T.R.~Taylor and B.~Zhu, \emph{{Differential equations for Carrollian amplitudes}}, \href{https://doi.org/10.1007/JHEP09(2024)149}{\emph{JHEP} {\bfseries 09} (2024) 149} [\href{https://arxiv.org/abs/2407.04789}{{\ttfamily 2407.04789}}].

\bibitem{Ciambelli:2023tzb}
L.~Ciambelli and D.~Grumiller, \emph{{Carroll geodesics}},  \href{https://arxiv.org/abs/2311.04112}{{\ttfamily 2311.04112}}.

\bibitem{Tadros:2024bev}
P.~Tadros and I.~Kol\'a\v{r}, \emph{{Carroll black holes in (A)dS and their higher-derivative modifications}},  \href{https://arxiv.org/abs/2408.01836}{{\ttfamily 2408.01836}}.

\bibitem{Utiyama:1956sy}
R.~Utiyama, \emph{{Invariant theoretical interpretation of interaction}}, \href{https://doi.org/10.1103/PhysRev.101.1597}{\emph{Phys. Rev.} {\bfseries 101} (1956) 1597}.

\bibitem{Sciama:1962}
D.W.~{Sciama}, \emph{{On the analogy between charge and spin in general relativity}},  in \emph{Recent Developments in General Relativity}, p.~415 (1962).

\bibitem{Kibble:1961ba}
T.W.B.~Kibble, \emph{{Lorentz invariance and the gravitational field}}, \href{https://doi.org/10.1063/1.1703702}{\emph{J. Math. Phys.} {\bfseries 2} (1961) 212}.

\bibitem{Duval:2014uoa}
C.~Duval, G.W.~Gibbons, P.A.~Horvathy and P.M.~Zhang, \emph{{Carroll versus Newton and Galilei: two dual non-Einsteinian concepts of time}}, \href{https://doi.org/10.1088/0264-9381/31/8/085016}{\emph{Class. Quant. Grav.} {\bfseries 31} (2014) 085016} [\href{https://arxiv.org/abs/1402.0657}{{\ttfamily 1402.0657}}].

\bibitem{Ciambelli:2019lap}
L.~Ciambelli, R.G.~Leigh, C.~Marteau and P.M.~Petropoulos, \emph{{Carroll Structures, Null Geometry and Conformal Isometries}}, \href{https://doi.org/10.1103/PhysRevD.100.046010}{\emph{Phys. Rev. D} {\bfseries 100} (2019) 046010} [\href{https://arxiv.org/abs/1905.02221}{{\ttfamily 1905.02221}}].

\bibitem{Herfray:2021qmp}
Y.~Herfray, \emph{{Carrollian manifolds and null infinity: a view from Cartan geometry}}, \href{https://doi.org/10.1088/1361-6382/ac635f}{\emph{Class. Quant. Grav.} {\bfseries 39} (2022) 215005} [\href{https://arxiv.org/abs/2112.09048}{{\ttfamily 2112.09048}}].

\bibitem{Figueroa-OFarrill:2020gpr}
J.~Figueroa-O'Farrill, \emph{{On the intrinsic torsion of spacetime structures}},  \href{https://arxiv.org/abs/2009.01948}{{\ttfamily 2009.01948}}.

\bibitem{Casalbuoni:2023bbh}
R.~Casalbuoni, D.~Dominici and J.~Gomis, \emph{{Two interacting conformal Carroll particles}}, \href{https://doi.org/10.1103/PhysRevD.108.086005}{\emph{Phys. Rev. D} {\bfseries 108} (2023) 086005} [\href{https://arxiv.org/abs/2306.02614}{{\ttfamily 2306.02614}}].

\bibitem{Ciambelli:2023mir}
L.~Ciambelli, L.~Freidel and R.G.~Leigh, \emph{{Null Raychaudhuri: canonical structure and the dressing time}}, \href{https://doi.org/10.1007/JHEP01(2024)166}{\emph{JHEP} {\bfseries 01} (2024) 166} [\href{https://arxiv.org/abs/2309.03932}{{\ttfamily 2309.03932}}].

\bibitem{Freidel:2024emv}
L.~Freidel and P.~Jai-akson, \emph{{Geometry of Carrollian Stretched Horizons}},  \href{https://arxiv.org/abs/2406.06709}{{\ttfamily 2406.06709}}.

\end{thebibliography}\endgroup
\end{document}